\documentclass[reprint,amsmath,amssymb,aps,prb,twocolumn,superscriptaddress]{revtex4-2}
\usepackage{graphicx}
\usepackage{amsmath}
\usepackage{amssymb}
\usepackage{tikz}
\usepackage{color} 
\usepackage{comment}
\usepackage{caption}
\usepackage{subcaption}
\usepackage{mathrsfs}

\usepackage{hyperref}
\hypersetup{colorlinks=true, linkcolor=blue, citecolor=black,urlcolor=blue}

\begin{document}

\title{Multistability by Design in Complex Triangular Mechanical Metamaterials}

\author{Chaviva Sirote-Katz}
\email{chavivas@mail.tau.ac.il}
\affiliation{School of Biomedical Engineering, Tel Aviv University, Tel Aviv 69978, Israel}

\author{Yair Shokef}
\email{shokef@tau.ac.il}
\affiliation{School of Mechanical Engineering, Tel Aviv University, Tel Aviv 69978, Israel}
\affiliation{School of Physics and Astronomy, Tel Aviv University, Tel Aviv 69978, Israel}
\affiliation{Center for Computational Molecular and Materials Science, Tel Aviv University, Tel Aviv 69978, Israel}
\affiliation{Center for Physics and Chemistry of Living Systems, Tel Aviv University, 69978, Tel Aviv, Israel}
\affiliation{International Institute for Sustainability with Knotted Chiral Meta Matter (WPI-SKCM$^2$), Hiroshima University, Higashi-Hiroshima, Hiroshima 739-8526, Japan}

\begin{abstract}
We introduce frustrated triangular networks that mix two different beam thicknesses. The two thicknesses separate the energetic costs of second-order buckling and angular deformation into four competing contributions. Using scaling arguments and mapping to an effective Ising description, we construct the system's phase diagram in terms of its dimensionless geometric parameters. Selected beam arrangements exhibit local bistability: hexagonal motifs switch independently between opposite twisting states, while linear motifs support independently switchable beam states. In both cases, the number of mechanically stable configurations grows exponentially with system size. Experiments on fabricated silicone metamaterials confirm the predicted bistability and local switching. More generally, the allowed beam arrangements map onto rhombus tilings, producing a large combinatorial space of architectures. The system thus combines multiplicity in both architecture and stable deformation states, establishing beam-thickness patterning as a route for programming frustration, multistability, and extensive degeneracy in triangular mechanical metamaterials.
\end{abstract}

\maketitle

\section{Introduction}

Mechanical metamaterials obtain their functionality primarily from architecture rather than solely from the properties of the constituent material.~\cite{bertoldi2017flexible, yu2018mechanical,Barchiesi2019,Lu2022,dudek2025shape}. Buckling instabilities, in particular, have been used to program deformation patterns, symmetry breaking, multistability, and other unusual mechanical responses~\cite{kang2013buckling,janbaz2016geometry,shan2015multistable}. Their geometry can therefore be designed to produce mechanical responses that are difficult to obtain in conventional materials, including auxetic behavior \cite{bertoldi2010negative}, spatially textured deformation~\cite{coulais2016combinatorial,sirote2025breaking}, and mechanical memory~\cite{kwakernaak2023counting,sirote2024emergent,meulblok2026transients,meulblok2026path}. More broadly, architectural design has been used to control buckling and instability patterns~\cite{kang2013buckling,janbaz2016geometry,findeisen2017characteristics}, generate multistability and shape retention~\cite{shan2015multistable,oppenheimer2015shapeable}, and produce unusual shape changes such as negative swelling~\cite{liu2016harnessing}. Similar principles have also been applied to hierarchical and bio-inspired networks~\cite{xu2015soft,ma2016nonlinear}, responsive fiber networks~\cite{sharabani2022messy,sharabani2024directional}, and actively reconfigurable pneumatic~lattices \cite{yuan2017tunable,zhu2026pneumatic}. Of particular interest are systems in which the preferred local deformations cannot all be simultaneously satisfied. This geometric frustration can produce collective deformation patterns and multiple mechanically stable states~\cite{sirote2024emergent}. When a mechanical metamaterial constructed from elastic beams is externally compressed, each beam preferentially buckles in its first-order, symmetric mode while preserving the angles between neighboring beams. For beams arranged to form a triangular lattice, however, these local preferences are mutually incompatible, as seen in Fig.~\ref{fig:intro}(a). The resulting geometric frustration is analogous to frustration in spin systems, where locally preferred interactions cannot all be simultaneously satisfied~\cite{Wannier1950, Han2008,shokef_PNAS_2011,ronceray2019range}. Related forms of frustration in mechanical metamaterials have been shown to generate defect-controlled responses, multistability, and unconventional global order~\cite{meeussen2020topological,udani2022taming,guo2023nonorientable}.

A uniform triangular beam network resolves this frustration differently depending on the beam thickness. For slender beams, collective changes of the angles between neighboring beams are energetically favorable. For thicker beams, changing these angles becomes more costly and individual beams instead buckle in their second-order mode. The crossover occurs at a normalized beam thickness $\hat{t}=t/L$ of approximately $\hat{t}_c\simeq0.235$, where $t$ is the beam thickness and $L$ is the lattice constant. Interestingly, the competition between these deformation mechanisms produces a transition between two ordered states, but does not give rise to extensive ground-state degeneracy~\cite{kang}.

In this paper, we introduce additional richness by allowing two different beam thicknesses within the same triangular lattice. We refer to beams of thickness $t_b$ as \emph{blue beams} and to beams of thickness $t_p=\alpha t_b$ as \emph{pink beams}, and impose a local constraint that every triangle contains exactly one pink beam. Thus, for $\alpha>1$, each triangle contains one thick and two thin beams, whereas for $\alpha<1$, it contains two thick and one thin beam. Even this simple constraint allows multiple different spatial arrangements of the blue and pink beams. Related combinatorial approaches have recently been used to design large families of floppy modes and frustrated loops and to obtain many programmable acoustic properties~\cite{liu2026combinatorial,keogh2025combinatorial}. As we show below, the arrangements of blue and pink beams map directly onto the classical problem of rhombus tilings. Beyond the combinatorial multiplicity of possible metamaterials that may be constructed from these simple anisotropic units, our approach allows to restore frustration such that the different resulting structures can each have multiple possible textured deformation states upon global compression.

The introduction of two beam thicknesses changes the energetic competition. The energetic cost of second-order buckling is now different for pink and blue beams, while angular deformation at a blue--blue junction has a different cost from angular deformation at a blue--pink junction. Constraining each triangle to have exactly one pink beam, we avoid the additional minor complexity of having also pink--pink junctions. The single energetic competition of the uniform lattice therefore separates into four contributions. Using scaling arguments and an effective Ising description, we compare these contributions and analytically construct a phase diagram, which is controlled by the normalized beam thickness and the ratio between the two beam thicknesses.

We first use a periodic hexagonal arrangement of pink beams to identify the deformation associated with the different regions of this phase diagram. Within one of these regimes, the hexagons themselves become bistable and can be switched independently, producing an extensive number of mechanically stable states. We then return to the much larger set of pink-beam arrangements allowed by the local constraint. Although a complete phase diagram for an arbitrary arrangement is considerably more complicated, the energetic hierarchy provides a way to understand the local deformation patterns that appear. In particular, line motifs occurring within these structures become locally bistable in another region of the phase diagram. Our experiments on fabricated metamaterials demonstrate the local switching of the hexagonal motifs.

\section{Model and Energetic Description}

\begin{figure}[ht]
\centering
\begin{subfigure}{0.48\textwidth}
\centering
\includegraphics[width=0.77\linewidth]{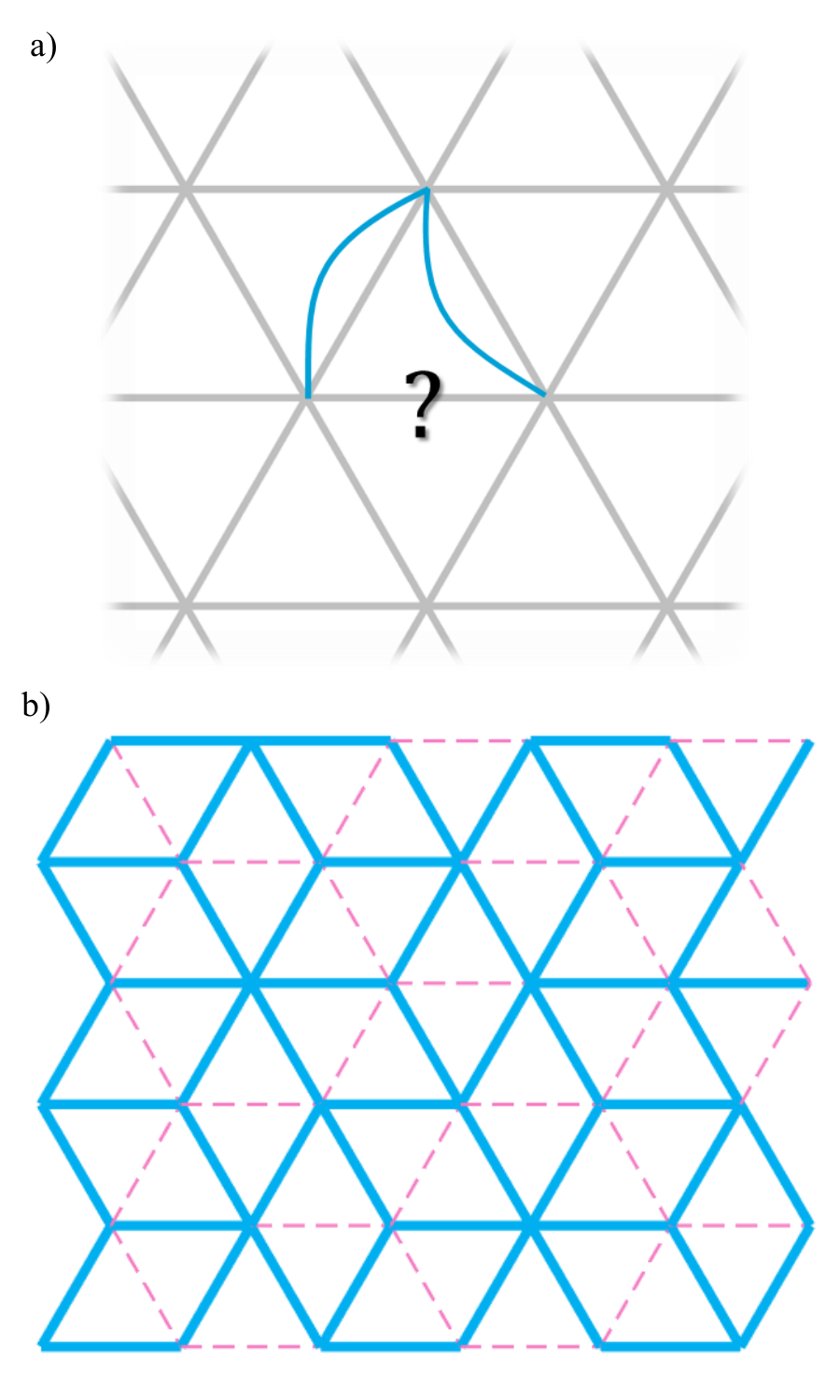}
\end{subfigure}
\begin{subfigure}{0.48\textwidth}
\centering
\includegraphics[width=0.77\linewidth]{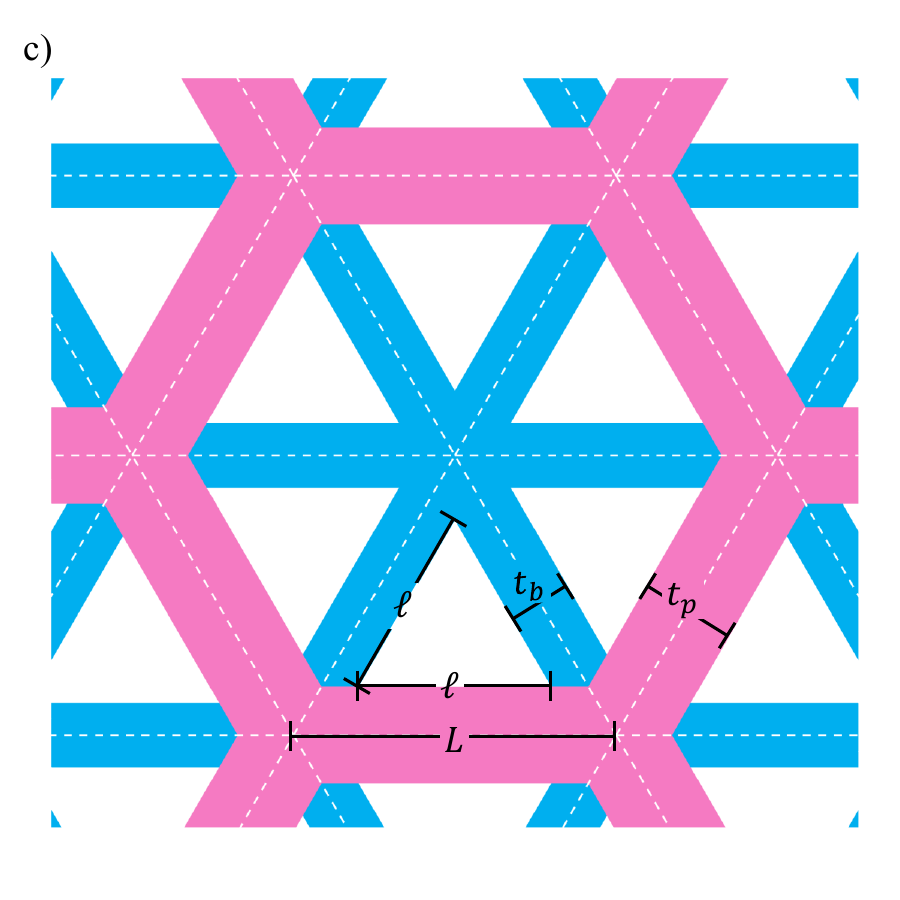}   
\end{subfigure}
\caption{\textbf{Combinatorial triangular mechanical metamaterial.} 
(a)~Geometric frustration of first-order beam buckling. 
(b)~One pink beam per triangle allows to design non-periodic metamaterials; removing the pink beams gives the corresponding rhombus tiling. 
(c)~Ordered hexagonal pink-beam pattern and geometric parameters $t_b$, $t_p$, $L$, and $\ell$.}
\label{fig:intro}
\end{figure}

We consider a triangular lattice with lattice constant~$L$ composed of beams of two different thicknesses. Blue beams have thickness~$t_b$, while pink beams have thickness $t_p=\alpha t_b$. In this case, we define $\hat{t}=t_b/L$ as the normalized blue-beam thickness, such that $\alpha$ and $\hat{t}$ provide two independent dimensionless geometric control parameters. We impose a constraint that every triangular unit contains exactly one pink beam and two blue beams. This local rule does not specify a unique structure. Instead, the orientation of the pink beam can vary between triangles while maintaining this constraint.

The number of possible structures can be understood by mapping the pink-beam arrangements onto rhombus tilings. Every pink beam is shared by the two triangles on either side of it. Pairing these triangles across their common pink beam forms a rhombus. Since every triangle contains exactly one pink beam, every triangle belongs to exactly one such pair. A valid pink-beam arrangement therefore defines a tiling of the triangular-lattice domain by rhombi, as seen in Fig.~\ref{fig:intro}(b). Conversely, a rhombus tiling uniquely specifies the pink beams by identifying the shared edge of each paired set of triangles.

Thus, asking how many different pink-beam orientations satisfy the one-pink-beam-per-triangle constraint is equivalent to asking how many rhombus tilings exist for a finite domain of the triangular lattice. For a hexagonal domain whose side lengths, measured in units of the triangular-lattice spacing, are $a,b,c,a,b,c$ in cyclic order, the number of such tilings is given by~\cite{MacMahon1915},
\begin{equation}
\Omega =\prod_{i=1}^{a}\prod_{j=1}^{b}\prod_{k=1}^{c}
\frac{i+j+k-1}{i+j+k-2}.
\label{eq:tilings}
\end{equation}
This can equivalently be expressed in terms of
hyperfactorials, $\mathscr{F}(n)=\prod_{k=0}^{n-1}k!$~\cite{kuperberg1994symmetries},
\begin{equation}
\Omega = \frac{\mathscr{F}(a)\mathscr{F}(b)\mathscr{F}(c)\mathscr{F}(a+b+c)}{\mathscr{F}(a+b)\mathscr{F}(b+c)\mathscr{F}(c+a)} .
\end{equation}
For the symmetric case $a=b=c=H$, the region contains
$N=6H^2$ unit triangles, where $H$ is its side length in units of
the triangular lattice spacing. The number of such tilings reduces to 
\begin{equation}
\Omega(H)=\frac{\mathscr{F}(H)^3\mathscr{F}(3H)}{\mathscr{F}(2H)^3} .
\label{eq:tilings_symmetric}
\end{equation}
For large $n$, $\log\mathscr{F}(n) \approx \frac{n^2}{2}\log n$~\cite{Ferreira2001DoubleGamma}. Substituting this expansion into Eq.~\eqref{eq:tilings_symmetric} gives
\begin{equation}
\log\Omega(H) \approx \left(\frac{9}{2}\log 3-6\log 2\right)H^2 \approx 0.785H^2 \approx 0.131N.
\end{equation}
Thus, the number of tilings grows exponentially with the number of triangles in the domain. The local thickness constraint therefore generates an exponentially large combinatorial space of possible metamaterial architectures. This connection between local architectural choices and a large space of possible global structures is related to broader combinatorial and inverse-design approaches in mechanical and acoustic metamaterials~\cite{coulais2016combinatorial,ronellenfitsch2019inverse,dieleman2020jigsaw,pisanty2020putting, sirote2025breaking,liu2026combinatorial,keogh2025combinatorial}. In these approaches, combinations of discrete local elements are used to encode target deformation modes, frustrated loops, spectral properties, or wave responses.

We next ask how our introduction of two beam thicknesses changes the mechanical response of these structures. As in the uniform triangular lattice, frustration may be resolved either by changing the angles between neighboring beams or by allowing individual beams to buckle in their second-order mode. With two beam thicknesses, however, each of these mechanisms separates into two distinct energetic contributions.

We consider the system under a prescribed external compression. In the absence of geometric frustration, each beam would buckle in its first mode with an amplitude determined by the imposed compression. We take the energy of this unfrustrated deformation as a baseline, such that the energetic contributions introduced below represent the additional costs arising from frustration.

We denote by $\Delta_b$ and $\Delta_p$ the energetic costs associated with second-order buckling of a blue and a pink beam, respectively. Similarly, $J_b$ denotes the energetic cost associated with changing the angle between two adjacent blue beams, while $J_x$ denotes the cost associated with changing the angle between a blue and a pink beam.

\begin{figure}[b]
\centering
\includegraphics[width=\columnwidth]{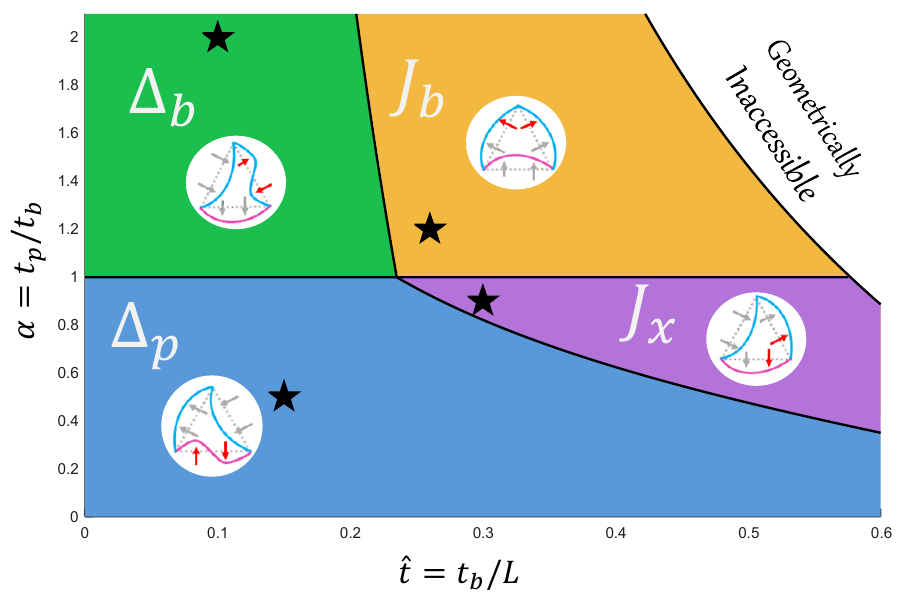}
\caption{\textbf{Phase diagram of the four dominant energetic regimes.}
The dominant energetic contribution is shown as a function of the normalized
beam thickness $\hat{t}$ and the thickness ratio $\alpha$. The colored regions
correspond to the four regimes $\Delta_b$, $\Delta_p$, $J_b$, and $J_x$, while
the white region is geometrically inaccessible. Insets show the deformation
of a single triangle that minimizes the corresponding energetic contribution.
The red arrows identify the spin pair associated with each
interaction: $\Delta_b$ couples the spins belonging to the same
blue beam, $\Delta_p$ couples the spins belonging to the same pink
beam, $J_b$ couples the neighboring spins on two blue beams meeting
at a vertex, and $J_x$ couples the neighboring spins on a blue and
a pink beam meeting at a vertex.
Black dots indicate the parameter values of the experimental samples.}
\label{fig:phase}
\end{figure}

Following the Ising description of a triangular lattice with uniform beam thickness~\cite{kang}, we assign two spins to each beam, one at each of its ends, to encode its buckling mode, see insets in Fig.~\ref{fig:phase}. Each spin takes the value $s_i=\pm1$ and represents the direction of the transverse displacement at the corresponding beam end. For each of the three beam orientations in the triangular lattice, we define a fixed positive transverse direction, such that the sign of $s_i$ is defined consistently throughout the lattice. When the two spins belonging to the same beam have the same sign, the beam buckles in its first mode, whereas opposite signs correspond to buckling in its second mode.

The total energy may then be written as
\begin{equation}
\begin{split}
\mathcal{H}={}&-\frac{\Delta_b}{2}
\sum_{\langle i,j\rangle_b}s_i s_j
-\frac{\Delta_p}{2}
\sum_{\langle i,j\rangle_p}s_i s_j\\
&-J_b\sum_{\langle i,j\rangle_{bb}}s_i s_j
-J_x\sum_{\langle i,j\rangle_{bp}}s_i s_j
+J_2\sum_{\mathrm{NNN}}s_i s_j .
\end{split}
\label{eq:Hamiltonian}
\end{equation}
Here, the first two sums run over the pairs of spins belonging to the same blue and pink beam, respectively. The corresponding energy scales, $\Delta_b$ and~$\Delta_p$, therefore determine the energetic preference of each beam to be in its first rather than second buckling mode. The third sum runs over neighboring spins belonging to two adjacent blue beams that meet at a vertex, with coupling strength~$J_b$, while the fourth runs over neighboring spins belonging to an adjacent blue and pink beam that meet at a vertex, with coupling strength~$J_x$. These terms describe the local interaction between the buckling directions of adjacent beams. The spin pairs corresponding to the four dominant energetic contributions are indicated in the insets of Fig.~\ref{fig:phase}. 

The final term runs over next-nearest-neighbor spin pairs and introduces a weaker, longer-range interaction with coupling strength~$J_2$. We expect this longer-range interaction to be weak compared with the dominant energetic scales,
\begin{equation}
J_2 \ll J_b,J_x,\Delta_b,\Delta_p,
\end{equation}
such that $J_2$ does not determine the primary energetic regimes. As we discuss below, its role instead becomes important in lifting degeneracies between configurations that are equivalent at the level of the dominant interactions. 

To estimate the dependence of the four dominant energies on beam geometry, we use Euler--Bernoulli beam theory. The energetic cost associated with second-order buckling of a beam scales approximately as~\cite{TimoshenkoGere1961}
\begin{equation}
    \Delta\sim\frac{\epsilon E w t^3}{\ell},
    \label{eq:delta}
\end{equation}
where $E$ is the material's Young's modulus, $\epsilon$ is the applied strain, $w$ is the out-of-plane width, $t$ is the beam thickness, and $\ell$ is the free length of the beam.

The energetic cost associated with an angular deformation at a joint scales as~\cite{kang}
\begin{equation}
    J\sim\epsilon E w t^2,
    \label{eq:J}
\end{equation}
where the factor $t^2$ arises from the product of the thicknesses of the two beams meeting at the joint. For a lattice with uniform beam thickness (i.e. $\alpha=1$), both beams have thickness~$t$, giving $t\times t=t^2$. The ratio between the two energies therefore scales as
\begin{equation}
    \frac{\Delta}{J}\sim\frac{t}{L},
\end{equation}
which gives the thickness-dependent competition observed in the uniform system.

For our lattice with two thicknesses of beams, $t_b$ and $t_p=\alpha t_b$, we obtain the two second-order buckling costs by applying the scaling of Eq.~(\ref{eq:delta}) separately to each beam type. Thus, $\Delta_b$ is evaluated using $t_b$, whereas $\Delta_p$ is evaluated using $t_p=\alpha t_b$. For the angular interactions, we assume that the energetic cost at a junction between beams of thicknesses $t_i$ and $t_j$ scales as the product $t_i t_j$. This reduces to the $t^2$ dependence of the uniform lattice when $t_i=t_j=t$. Therefore, $J_b$ scales with $t_b^2$ for a junction between two blue beams, while $J_x$ scales with $t_b t_p=\alpha t_b^2$ for a junction between a blue and a pink beam. The free beam lengths differ slightly from the lattice constant $L$ because the finite beam thickness modifies the geometry close to each junction, however since the triangular holes between the beams are equilateral (see Fig.~\ref{fig:intro}), all beams have the same free length,
\begin{equation}
    \ell=L-\frac{2+\alpha}{\sqrt{3}}t_b.
    \label{eq:L_vs_ell}
\end{equation}
Thus the four dominant energetic contributions scale with the dimensionless geometric parameters $\hat{t}$ and $\alpha$ as
\begin{align}
    \Delta_b &\sim
    \frac{t_b^3}{\ell} \sim \frac{\hat{t}^3} {1-\frac{2+\alpha}{\sqrt{3}}\hat{t}},
    \label{eq:Db}\\
    \Delta_p &\sim
    \frac{(\alpha t_b)^3}{\ell} \sim \frac{ \alpha^3\hat{t}^3} {1-\frac{2+\alpha}{\sqrt{3}}\hat{t}},
    \label{eq:Dp}\\
    J_b &\sim
    t_b^2 \sim \hat{t}^2,
    \label{eq:Jb}\\
    J_x &\sim
    (\alpha t_b)\cdot t_b \sim \alpha \hat{t}^2.
    \label{eq:Jx}
\end{align}

The different dependence of Eqs.~\eqref{eq:Db}--\eqref{eq:Jx} on~$\alpha$ and~$\hat{t}$ separates energetic contributions that are equivalent in the uniform lattice. This produces a two-dimensional competition between $\Delta_b$, $\Delta_p$, $J_b$, and $J_x$, which we consider next. For the subsequent comparison that we will perform between these four energies, we emphasize that we will need their scaling with $\hat{t}$ and with $\alpha$, and the facts that the prefactors in the expressions for $\Delta_b$ and $\Delta_p$ are equal, and similarly the prefactors for $J_b$ and $J_x$ are also equal.

\section{Phase Diagram and Energetically Preferred States}

We construct a phase diagram by comparing the four dominant energetic costs in the $(\alpha,\hat{t})$ plane. The diagram identifies the least costly local mechanism for accommodating frustration. For the periodic hexagonal architecture considered below, this hierarchy determines the corresponding minimal-energy deformation. For a general pink-beam arrangement, it should instead be interpreted as a local energetic guide rather than a complete global phase diagram. Each region of Fig.~\ref{fig:phase} indicates which of $\Delta_b$, $\Delta_p$, $J_b$, and $J_x$ is smallest, with the boundaries between neighboring regions analytically obtained by equating the corresponding energetic costs.

For the uniform system, $\alpha=1$, the transition occurs at a normalized critical thickness of $\hat{t}_c=t_c/L\simeq0.235$~\cite{kang}. Using Eq.~(\ref{eq:L_vs_ell}), we express this transition in terms of the free beam length as
\begin{equation}
    C=\frac{t_c}{\ell}
    =\frac{t_c/L}{1-\sqrt{3}t_c/L}
    \simeq 0.396.
\end{equation}

The line $\alpha=1$ is itself a phase boundary, corresponding to the crossover $\Delta_p=\Delta_b$ as well as $J_x=J_b$. For $\alpha>1$, where the pink beams are thicker than the blue beams, the remaining phase boundary is obtained by equating $\Delta_b$ and $J_b$ and by requiring that in the $\alpha=1$ limit, the transition will occur at $\hat{t}=\hat{t}_c$. This leads to
\begin{equation}
    \frac{\hat{t}}
    {1-\frac{2+\alpha}{\sqrt{3}}\hat{t}}=C,
\end{equation}
or equivalently,
\begin{equation}
    \hat{t}=
    \frac{\sqrt{3}C}
    {\sqrt{3}+C(2+\alpha)}.
\end{equation}
For $\alpha<1$, where the pink beams are thinner than the blue beams, the corresponding phase boundary, obtained from equating $\Delta_p$ and $J_x$ is
\begin{equation}
    \hat{t}=
    \frac{\sqrt{3}C}
    {\sqrt{3}\alpha^2+C(2+\alpha)}.
\end{equation}

\begin{figure}[h]
\centering
\includegraphics[width=1\columnwidth]{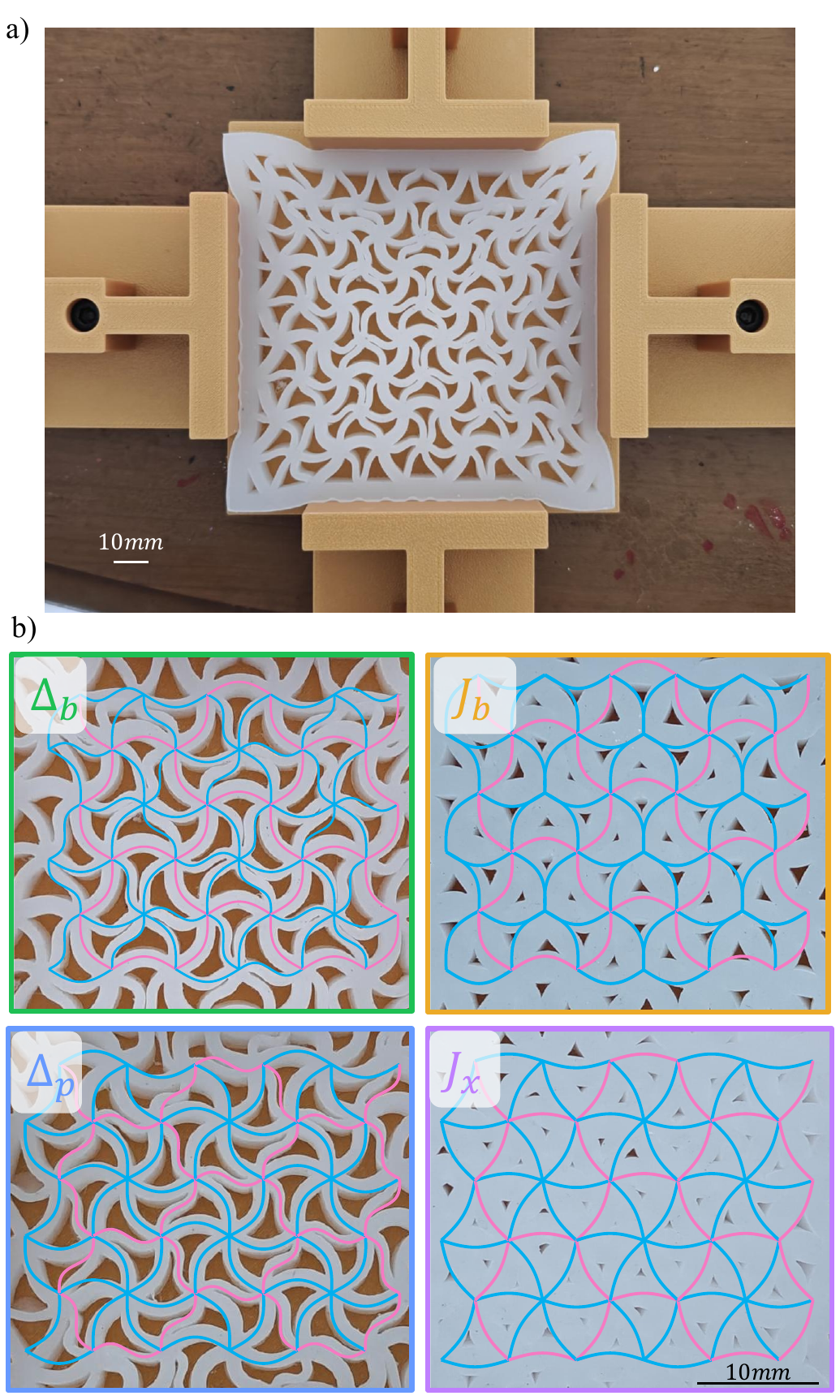}
\caption {\textbf{Theoretical and experimental realization of the four phases for hexagonal pink-bar patterns.} 
(a)~Custom-built biaxial compression setup used to compress the fabricated samples simultaneously along two perpendicular directions. (b)~Experimentally fabricated metamaterials with the theoretical predictions overlaid in pink and blue lines, showing the deformation patterns corresponding to the four dominant energetic contributions, as indicated on each image. The 3D-printed samples for $\Delta_b$, $\Delta_p$, $J_b$, and $J_x$ correspond to experimental parameters of $(\hat{t},\alpha)=(0.1,2)$, $(0.15,0.5)$, $(0.26,1.2)$, and $(0.3,0.9)$, respectively.}
\label{fig:hex}
\end{figure}

To connect the energetic phase diagram to actual deformation patterns, we first consider the periodic pink-beam arrangement shown in Fig.~\ref{fig:intro}, in which the pink beams form hexagons throughout the lattice. We fabricated samples by three-dimensional printing molds, casting silicone 00-30 into them, and subjecting the resulting structures to biaxial compression. The samples have a lattice spacing of $L=13mm$ and contain $N=16\times9=144$ triangular units. We tested samples corresponding to each of the four regions of the phase diagram and found excellent agreement between the experimentally observed deformation patterns and the theoretical predictions presented here, see Fig.~\ref{fig:hex}.

In the $\Delta_b$ regime, second-order buckling of blue beams is least costly, and the system preferentially accommodates its frustration through these beams. In the $\Delta_p$ regime, second-order buckling of the pink beams is favored instead. In the $J_b$ regime, angular changes between neighboring blue beams provide the least expensive mechanism, whereas in the $J_x$ regime the deformation preferentially involves angle changes at blue--pink junctions.

At $\alpha=1$, the phase diagram reduces to that of the uniform triangular lattice. Moving away from $\alpha=1$ introduces a distinction between the two beam thicknesses. For the hexagonal pattern considered here, this does not introduce a new deformation pattern, but instead selects between different translations of the same pattern.

Interestingly, one of these regimes contains an additional feature. When $\Delta_b$ is the lowest energetic cost, the hexagonal pattern does not have only one mechanically stable realization. Instead, each hexagonal motif can adopt two distinct configurations. We consider this local bistability in the following section.

\section{Multistability of the Hexagonal Pattern}

\begin{figure}[h!]
\centering
\includegraphics[width=0.95\columnwidth]{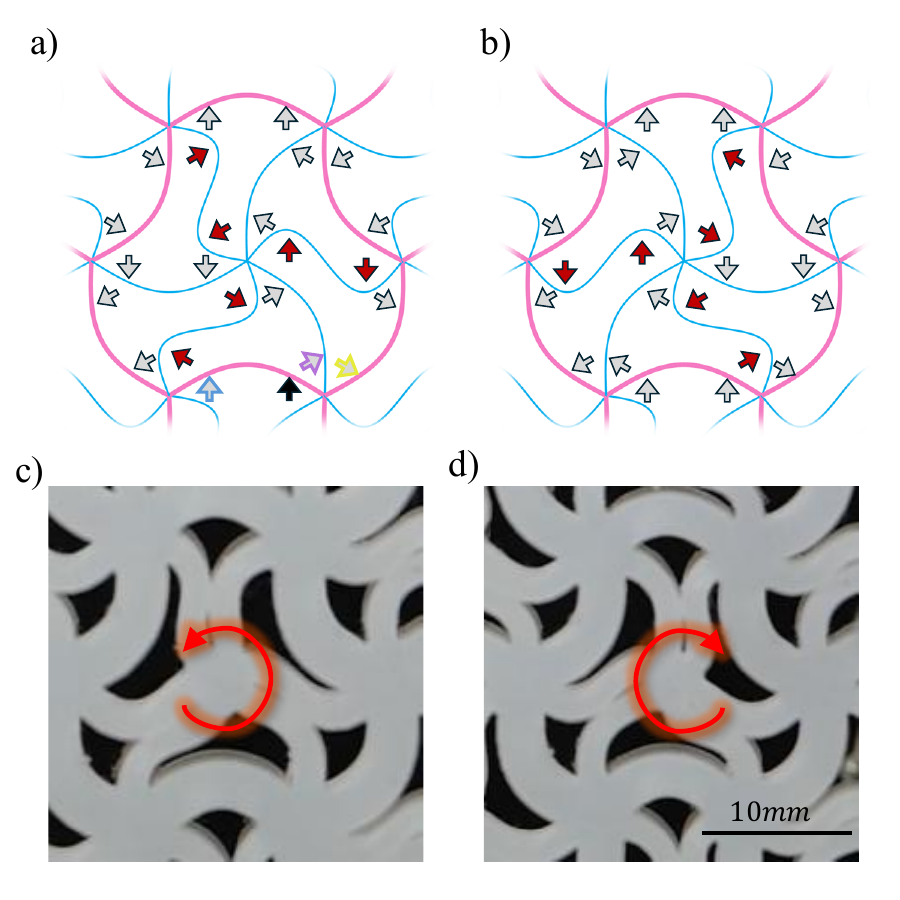}
\caption{ \textbf{Bistability of the hexagonal pattern}. 
(a,b)~Two degenerate spin configurations corresponding to opposite deformations of the central hexagon. Arrows indicate spins, with red arrows marking frustrated spin pairs. Purple, blue, and yellow distinguish the three spin interactions relative to the black reference spin, $J_x$, $\Delta_p$, $J_2$, respectively. 
(c,d)~3D-printed sample demonstrating the two stable states; The central hexagon is externally rotated from (c) to (d) and after release remains stable in the switched configuration.}
\label{fig:hex bi}
\end{figure}

We now focus on the $\Delta_b$ region of the phase diagram, where second-order buckling of a blue beam is the least costly way to accommodate frustration. The $\Delta_p$, $J_x$, and $J_2$ interactions are explicitly marked in Fig.~\ref{fig:hex bi}(a), where they are shown relative to a reference spin. Within a region enclosed by pink beams, the deformation can be represented by an alternating sequence of first- and second-order buckling modes. The sequences $1,2,1,2,\ldots$ and $2,1,2,1,\ldots$ contain the same number of costly deformations and are therefore energetically equivalent within this description.

For the hexagonal pink-beam pattern, these two sequences correspond to two opposite deformations of each hexagon, as shown in Fig.~\ref{fig:hex bi}. A hexagon may twist in one direction or in the opposite direction without changing the leading-order energetic cost. Moreover, changing the state of one hexagon does not require the surrounding hexagons to change their states.

The hexagons therefore act as independent local bistable elements, analogous to mechanical hysterons and to bistable beam elements used in other multistable architected materials~\cite{shan2015multistable,ding2022sequential}. For a structure containing only hexagons, the number of such stable states is
\begin{equation}
    \Omega_{\Delta_b}=2^{N/6} ,
\end{equation}
where $N$ is the number of triangles in the lattice.
Because $\log\Omega_{\Delta_b}$ is proportional to the number of triangular units~$N$, the corresponding configurational entropy is extensive.

To test the predicted local bistability, we used a fine-tipped probe to displace the beams in a single hexagon in the direction of the alternative twisted configuration and then released it. In the $\Delta_b$ regime, the manipulated hexagon remained in its new configuration after release rather than returning to its original state Fig.~\ref{fig:hex bi}(c,d). We repeated this procedure on different hexagons and experimentally demonstrated that they could be switched independently without changing the states of the surrounding hexagons. By contrast, when we performed analogous local manipulations on samples in the other three regimes of the phase diagram, the structures returned to their original configurations after release. These experiments directly demonstrate the predicted independent local bistability in the $\Delta_b$ regime.

\section{General Pink-Beam Patterns and Their Deformation}

The hexagonal pattern of blue and pink beams provides a particularly simple arrangement for connecting the energetic competition to the possible deformation patterns and for producing extensive multistability through multiple bistable motifs within the lattice. For this periodic architecture, the phase diagram in Fig.~\ref{fig:phase} is exact within our effective model. The deformation patterns in the four regimes are closely related to those previously obtained for the uniform triangular lattice~\cite{kang}; the introduction of the pink beams primarily shifts the deformation pattern relative to the lattice. In the $\Delta_b$ regime, it additionally introduces the local bistability described above. The hexagonal pattern represents, however, only one of the many possible pink-beam arrangements. The mapping to rhombus tilings shows that the one-pink-beam-per-triangle constraint permits a huge number of different pink-beam orientations. For these more general structures, the phase diagram should be interpreted as a local energetic guide rather than as a complete prediction of the deformations that will form across the entire lattice. We now consider how these structures deform within each of the four dominant energetic regimes.

The four energetic costs derived above are local and do not depend on the pink beams forming a periodic hexagonal pattern. We therefore expect that, at a point in the $(\alpha,\hat{t})$ plane where one of the four costs is substantially lower than the others, the system will preferentially accommodate frustration through the corresponding deformation mechanism. Thus, even for a complicated pink-beam arrangement, regions of the structure should preferentially exhibit second-order buckling of blue beams when $\Delta_b$ is smallest, second-order buckling of pink beams when $\Delta_p$ is smallest, blue--blue angle changes when $J_b$ is smallest, and blue--pink angle changes when $J_x$ is smallest.

Deep within each region in the $(\alpha,\hat{t})$ phase diagram, one energetic cost is substantially smaller than the other three and therefore determines the preferred deformation, resulting in one of the four phases described above. Near the boundaries between these regions, the distinction between the dominant energetic contributions becomes less clear, and the resulting behavior remains to be explored. We therefore restrict our analysis to the behavior well within each of the four dominant regimes. For this reason, the
phase diagram of Fig.~\ref{fig:phase} should be interpreted as identifying the
locally preferred deformation mechanism rather than as a complete prediction
of the global ground state for every possible pink-beam arrangement.
Nevertheless, for any fixed orientation of the pink beams, we can
construct deformation patterns corresponding to each of the four limiting
energetic regimes, as described below.

\begin{figure}[b]
\centering
\includegraphics[width=0.95\columnwidth]{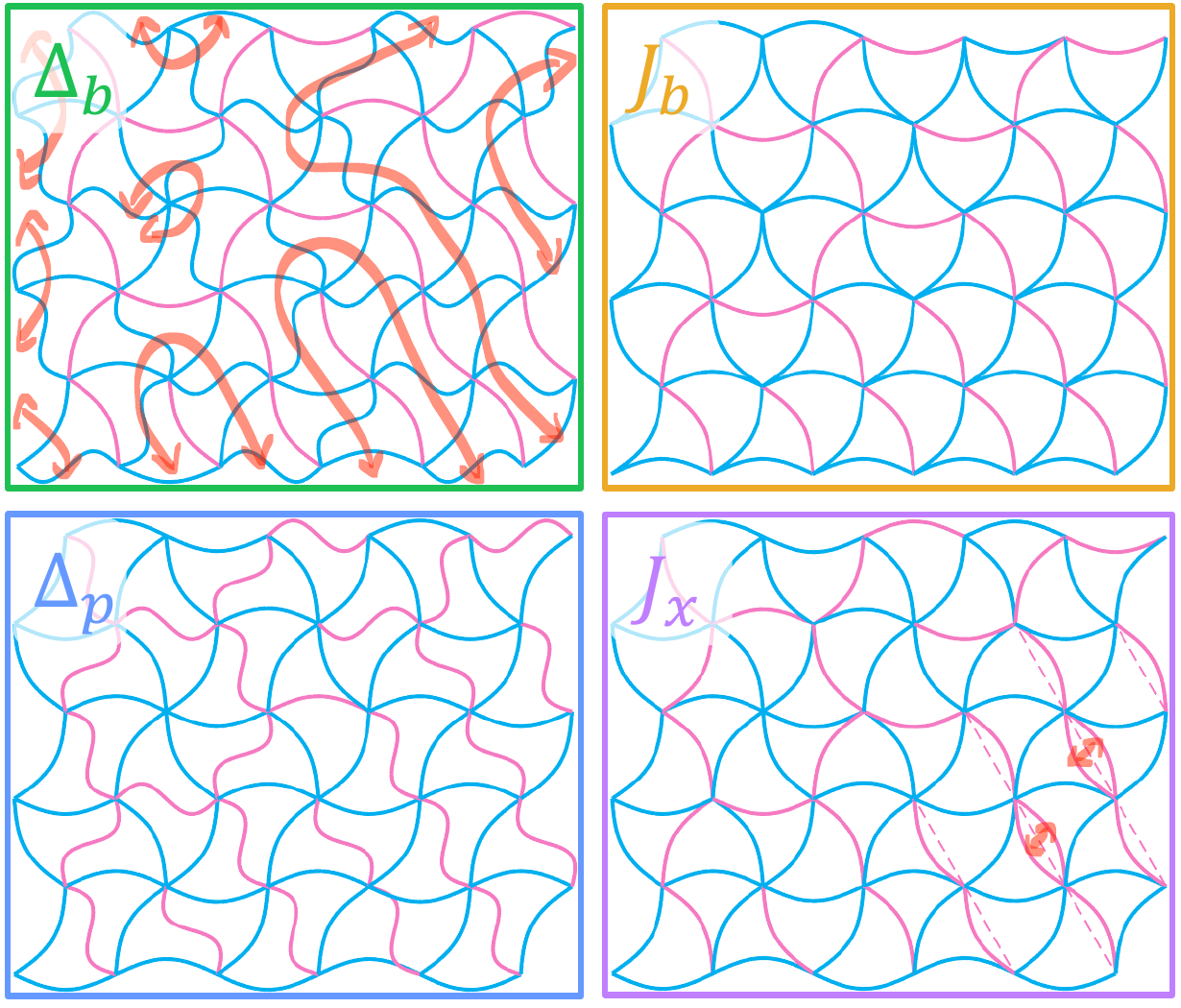}
\caption{\textbf{Deformation patterns in the four energetic regimes.}
Predicted deformation patterns of the triangular metamaterial deep into the $\Delta_b$, $J_b$, $\Delta_p$, and $J_x$ regimes. Blue and pink curves show the deformed beam configurations, while dashed pink lines indicate the undeformed states of the pink beams. Red double arrows indicate regions exhibiting bistability, where alternative deformation states are possible.}
\label{fig:complex}
\end{figure}

Figure~\ref{fig:complex} shows the deformation patterns obtained for a representative random arrangement of pink beams in each of the four energetic regimes. Despite the irregular arrangement of the pink beams, a corresponding deformation pattern can be constructed for each regime using a local, sequential procedure.

For any fixed arrangement of the pink beams, the pink beams divide the lattice into regions of connected blue beams. Each region consists of all the blue beams enclosed by a set of pink beams, as illustrated by the regions marked in red for the $\Delta_b$ phase in Fig.~\ref{fig:complex} (top left). The construction begins with one such blue-beam region. The deformation of its blue beams is assigned first, followed by the surrounding pink beams. The procedure then continues to a neighboring blue-beam region, whose deformation is chosen consistently with the beams that have already been assigned. At every step, the weaker next-nearest-neighbor interaction $J_2$ is satisfied whenever possible. This additional condition determines the orientations of the beams and ensures that the assignments in neighboring regions are consistent.

In the $\Delta_b$ regime, the blue beams within each region are assigned alternating first and second buckling modes. The surrounding pink beams are then added in deformations compatible with the blue beams without introducing further incompatibility. In the $\Delta_p$ regime, the blue beams are assigned first while satisfying $J_2$, after which the required pink beams are placed in their second buckling mode. The construction then proceeds to the next blue-beam region.

The same sequential procedure is used in the $J_b$ and $J_x$ regimes. In the $J_b$ regime, all the blue beams within a region are assigned to buckle in the same direction and in the same order. The surrounding pink beams are then added, and the next blue-beam region is assigned consistently with them while satisfying $J_2$ wherever possible. In the $J_x$ regime, the blue beams are assigned first, and each neighboring pink beam is given the deformation direction required to satisfy its interaction with the blue beam. The procedure is then repeated for the neighboring regions.

In the $\Delta_b$- and $\Delta_p$-dominated regimes, the dominant energetic contribution and the $J_2$ interactions can be satisfied simultaneously throughout the construction. In the $J_b$- and $J_x$-dominated regimes, some $J_2$ interactions may necessarily remain unsatisfied. Nevertheless, these local violations do not prevent the construction of a deformation pattern that satisfies the dominant energetic contribution for any arrangement of the pink beams.

A complete theoretical phase diagram for an arbitrary pink-beam pattern would require accounting for the interactions between its different motifs and for the many possible transition configurations between them. The number of possibilities increases rapidly with the complexity of the arrangement, and we do not attempt to enumerate them here. Instead, we use the energetic hierarchy to understand characteristic deformation patterns that emerge within more general structures.

An important example appears when the pink-beam arrangement contains a line motif. Consider three or more consecutive pink beams arranged along a line, as indicated by the dotted pink beams in Fig.~\ref{fig:complex} (bottom right). In the $J_x$ regime, where changing a blue--pink angle has the lowest energetic cost, the middle pink beam of such a motif can occupy two different configurations while maintaining the preferred deformation of its surroundings. The line motif therefore introduces a local bistability that is distinct from the hexagonal bistability discussed above.

By constructing an extended pattern containing such lines, multiple pink beams can become independently switchable. At the level of the four dominant energetic contributions, the alternative configurations are degenerate. Here, however, the weaker second-nearest-neighbor interaction $J_2$ must also be considered in order to distinguish the detailed deformation patterns, as indicated in Fig.~\ref{fig:complex}.

For an only line pattern containing $N$ triangles, there are $N/2$ independently switchable elements, the number of mechanically stable configurations is
\begin{equation}
    \Omega_{J_x}=2^{N/2}.
\end{equation}
Thus, extensive multistability is not restricted to a hexagonal pattern in region $\Delta_b$. It can also arise from particular local motifs embedded within the much larger combinatorial set of allowed pink-beam orientations.

Together, the hexagonal and line patterns show that different pink-beam arrangements can produce different forms of local multistability. The parameters $\alpha$ and $\hat{t}$ determine the energetic regime and therefore which deformation mechanisms are favored, while the arrangement of the pink beams determines which local motifs are present. The combination of these two design choices determines where independently switchable elements can occur within the metamaterial.

\section{Discussion}

Introducing two beam thicknesses into a frustrated triangular lattice provides two distinct levels of control over its mechanical response. First, the thicknesses determine the relative energetic costs of the available deformation mechanisms. Second, the constraint of one pink beam per triangle allows many different spatial arrangements of those thicknesses within the lattice.

The first level of control converts the single thickness-dependent competition of the uniform triangular lattice into a phase diagram involving four energetic contributions. Second-order buckling separates into $\Delta_b$ and $\Delta_p$ for blue and pink beams, respectively, while angular deformation separates into $J_b$ and $J_x$ for blue--blue and blue--pink junctions. Varying $\alpha$ and $\hat{t}$ therefore changes which local deformation mechanism is energetically preferred.

The second level of control is combinatorial. The mapping to rhombus tilings shows that the local one-pink-beam-per-triangle rule permits a large number of different architectures. These architectures are mechanically distinct because different orientations of the pink beams produce different local motifs and therefore different ways of accommodating frustration.

For the periodic hexagonal architecture, the energetic picture is sufficiently simple that we can directly relate the regions of the phase diagram to characteristic ground-state deformations. In the $\Delta_b$ regime, each pink-beam hexagon becomes a localized bistable element that can independently twist in either direction. A system containing $N_h$ such elements consequently has $2^{N_h}$ stable states, providing an explicit realization of extensive ground-state degeneracy.

More general pink-beam arrangements demonstrate that the same energetic framework can produce additional behavior. In particular, a line motif becomes locally bistable in the $J_x$ regime, and repeating this motif again produces exponentially many stable configurations. The two examples have different geometric origins: the $\Delta_b$ multistability is associated with the hexagonal pattern, whereas the $J_x$ multistability is associated with lines of consecutive pink beams. In both cases, the corresponding local elements are predicted to be independently switchable. For the hexagonal pattern, our experiments confirm that these elements can indeed be switched individually and remain in their new states after release.

For arbitrary pink-beam arrangements, predicting the complete global ground state remains more difficult. Although the lowest of the four energetic costs identifies the locally preferred deformation mechanism, the final deformation also depends on how different motifs meet and interact throughout the structure. The simple phase diagram therefore provides a guide to the local response rather than a complete theoretical phase diagram for every possible architecture. Developing a systematic description of these transition configurations would provide a route toward predicting the response of arbitrary patterns directly from their pink-beam orientations.

The combination of structural and mechanical multiplicity may be useful for designing mechanical memory and adaptive materials. A particular pink-beam arrangement determines where bistable motifs occur, while the beam thicknesses determine whether those motifs are mechanically active. Local bistable elements can consequently act as independently addressable mechanical states whose locations are encoded geometrically within the metamaterial. 

The ability to select and address particular deformation modes is also relevant to actively driven and reconfigurable metamaterials. Related systems use nonequilibrium activity to selectively actuate mechanical zero modes~\cite{woodhouse2018autonomous}, pneumatic control to modify lattice buckling and stiffness~\cite{yuan2017tunable,zhu2026pneumatic}, and local reorientation to reprogram acoustic and topological responses~\cite{keogh2025combinatorial,lemkalli2026nonreciprocal}. Although the structures studied here are passive, their large number of architectures and mechanically stable states could provide a basis for similar forms of externally controlled reconfiguration.

More broadly, these results show that a relatively small modification of a geometrically frustrated lattice can produce a much richer mechanical landscape. Starting from the simple rule that every triangle contains one pink beam, we obtain a combinatorial set of possible architectures. By tuning the two beam thicknesses, we then select between competing deformation mechanisms and identify architectures containing extensively many stable states. This provides a route for programming both the structure and the deformation landscape of frustrated triangular mechanical metamaterials.

\section*{Acknowledgments}
We thank Bat-El Pinchasik, Priyanka, Robin Selinger, Samudrajit Thapa, and Tomer Markovich for helpful discussions; Elisheva Berkowicz, Marcos (Moty) Dorfman, Ora Swidler, Rami Eliasi, Raziel Katz, Shai Sonnenreich, Yonathan Aharony, and Yotam Hantman for technical assistance; and Katia Bertoldi and Sung Hoon Kang for providing additional information on their work. The writing of this manuscript benefited from the use of ChatGPT (OpenAI, GPT-5.6) for grammar and wording improvements. C.S.K. was supported by the Clore Scholars Programme. This research was supported in part by the Israel Science Foundation Grant No. 1899/20.

\bibliographystyle{unsrt}
\bibliography{references}

@book{MacMahon1915,
  author    = {MacMahon, Percy Alexander},
  title     = {Combinatory Analysis},
  publisher = {Cambridge University Press},
  address   = {Cambridge},
  volume    = {2},
  year      = {1916}
}

@article{kang,
  title={Complex ordered patterns in mechanical instability induced geometrically frustrated triangular cellular structures},
  author={Kang, Sung Hoon and Shan, Sicong and Ko{\v{s}}mrlj, Andrej and Noorduin, Wim L and Shian, Samuel and Weaver, James C and Clarke, David R and Bertoldi, Katia},
  journal={Physical Review Letters},
  volume={112},
  number={9},
  pages={098701},
  year={2014},
  publisher={APS}
}

@article{sirote2025breaking,
  title={Breaking mechanical holography in combinatorial metamaterials},
  author={Sirote-Katz, Chaviva and Palti, Ofri and Spiro, Naomi and K{\'a}lm{\'a}n, Tam{\'a}s and Shokef, Yair},
  journal={Physical Review Research},
  volume={7},
  number={3},
  pages={033004},
  year={2025},
  publisher={APS}
}

@article{sirote2024emergent,
  title={Emergent disorder and mechanical memory in periodic metamaterials},
  author={Sirote-Katz, Chaviva and Shohat, Dor and Merrigan, Carl and Lahini, Yoav and Nisoli, Cristiano and Shokef, Yair},
  journal={Nature Communications},
  volume={15},
  number={1},
  pages={4008},
  year={2024},
  publisher={Nature Publishing Group UK London}
}

@article{coulais2016combinatorial,
  title={Combinatorial design of textured mechanical metamaterials},
  author={Coulais, Corentin and Teomy, Eial and De Reus, Koen and Shokef, Yair and Van Hecke, Martin},
  journal={Nature},
  volume={535},
  number={7613},
  pages={529--532},
  year={2016},
  publisher={Nature Publishing Group UK London}
}

@article{bertoldi2010negative,
  title   = {Negative Poisson's Ratio Behavior Induced by an Elastic Instability},
  author  = {Bertoldi, Katia and Reis, Pedro M. and Willshaw, Stephen and Mullin, Tom},
  journal = {Advanced Materials},
  volume  = {22},
  number  = {3},
  pages   = {361--366},
  year    = {2010},
  doi     = {10.1002/adma.200901956}
}

@article{bertoldi2017flexible,
  title={Flexible mechanical metamaterials},
  author={Bertoldi, Katia and Vitelli, Vincenzo and Christensen, Johan and Van Hecke, Martin},
  journal={Nature Reviews Materials},
  volume={2},
  number={11},
  pages={17066},
  year={2017},
  publisher={Nature Publishing Group}
}

@article{dudek2025shape,
  title={Shape-morphing metamaterials},
  author={Dudek, Krzysztof K and Kadic, Muamer and Coulais, Corentin and Bertoldi, Katia},
  journal={Nature Reviews Materials},
  volume={10},
  number={10},
  pages={783--798},
  year={2025},
  publisher={Nature Publishing Group UK London}
}

@book{TimoshenkoGere1961,
  author    = {Timoshenko, Stephen P. and Gere, James M.},
  title     = {Theory of Elastic Stability},
  edition   = {{Second}},
  publisher = {McGraw-Hill},
  address   = {New York},
  year      = {1961}
}

@article{Wannier1950,
  author  = {Wannier, Gregory H.},
  title   = {Antiferromagnetism. The Triangular {Ising} Net},
  journal = {Physical Review},
  volume  = {79},
  pages   = {357--364},
  year    = {1950},
  doi     = {10.1103/PhysRev.79.357}
}

@article{Han2008,
  author  = {Han, Yilong and Shokef, Yair and Alsayed, Ahmed M. and
             Yunker, Peter and Lubensky, Tom C. and Yodh, Arjun G.},
  title   = {Geometric frustration in buckled colloidal monolayers},
  journal = {Nature},
  volume  = {456},
  pages   = {898--903},
  year    = {2008},
  doi     = {10.1038/nature07595}
}

@article{shokef_PNAS_2011,
	title = {Order by disorder in the antiferromagnetic {Ising} model on an elastic triangular lattice},
	volume = {108},
	doi = {10.1073/pnas.1014915108},
	journal = {Proceedings of the National Academy of Sciences},
	author = {Shokef, Yair and Souslov, Anton and Lubensky, Tom C.},
	year = {2011},
	pages = {11804--11809},
}

@article{yu2018mechanical,
  title   = {Mechanical Metamaterials Associated with Stiffness, Rigidity and Compressibility: A Brief Review},
  author  = {Yu, Xianglong and Zhou, Ji and Liang, Haiyi and Jiang, Zhengyi and Wu, Lingling},
  journal = {Progress in Materials Science},
  volume  = {94},
  pages   = {114--173},
  year    = {2018},
  doi     = {10.1016/j.pmatsci.2017.12.003}
}

@article{Barchiesi2019,
  author  = {Barchiesi, Emilio and Spagnuolo, Mario and Placidi, Luca},
  title   = {Mechanical metamaterials: a state of the art},
  journal = {Mathematics and Mechanics of Solids},
  volume  = {24},
  number  = {1},
  pages   = {212--234},
  year    = {2019},
  doi     = {10.1177/1081286517735695}
}

@article{Lu2022,
  author  = {Lu, Chenxi and Hsieh, Mengting and Huang, Zhifeng and Zhang, Chi
             and Lin, Yaojun and Shen, Qiang and Chen, Fei and Zhang, Lianmeng},
  title   = {Architectural Design and Additive Manufacturing of Mechanical Metamaterials: A Review},
  journal = {Engineering},
  volume  = {17},
  pages   = {44--63},
  year    = {2022},
  doi     = {10.1016/j.eng.2021.12.023}
}

@article{ding2022sequential,
  title={Sequential snapping and pathways in a mechanical metamaterial},
  author={Ding, Jiangnan and van Hecke, Martin},
  journal={The Journal of Chemical Physics},
  volume={156},
  number={20},
  year={2022},
  publisher={AIP Publishing}
}

@article{liu2026combinatorial,
  title = {Combinatorial Design of Floppy Modes and Frustrated Loops in Metamaterials},
  author = {Liu, Wenfeng and Sigalov, Tomer A. and Coulais, Corentin and Shokef, Yair},
  journal = {Physical Review Letters},
  volume = {136},
  pages = {038202},
  year = {2026},
  doi = {10.1103/nqgr-tfb1}
}

@article{meeussen2020topological,
  title = {Topological Defects Produce Exotic Mechanics in Complex Metamaterials},
  author = {Meeussen, Anne S. and O{\u{g}}uz, Erdal C. and Shokef, Yair and van Hecke, Martin},
  journal = {Nature Physics},
  volume = {16},
  pages = {307--311},
  year = {2020},
  doi = {10.1038/s41567-019-0763-6}
}

@article{lemkalli2026nonreciprocal,
  title = {Nonreciprocal Topological Kink-Wave Propagation in Mechanical Metamaterials},
  author = {Lemkalli, Brahim and Ji, Qingxiang and Zhang, Jingyi and Craster, Richard V. and Christensen, Johan and Kadic, Muamer},
  year = {2026},
  journal = {arXiv:2602.04591},
  doi = {10.48550/arXiv.2602.04591}
}

@inproceedings{yuan2017tunable,
  title = {Tunable Triangular Cellular Structures by Pneumatic Control of Dual Channel Actuators},
  author = {Yuan, Zhihao and Ju, Jaehyung},
  booktitle = {Proceedings of the ASME 2017 International Mechanical Engineering Congress and Exposition},
  volume = {9},
  pages = {V009T12A013},
  year = {2017},
  publisher = {American Society of Mechanical Engineers},
  doi = {10.1115/IMECE2017-70858}
}

@article{zhu2026pneumatic,
  title = {Pneumatically Controlled Lattices with Tunable Mechanical Behavior},
  author = {Zhu, Xiaoheng and Hua, Yucong and Jin, Dengge and Raney, Jordan R.},
  journal = {Communications Engineering},
  volume = {5},
  pages = {14},
  year = {2026},
  doi = {10.1038/s44172-025-00570-8}
}

@article{janbaz2016geometry,
  title = {Geometry-Based Control of Instability Patterns in Cellular Soft Matter},
  author = {Janbaz, Shahram and Weinans, Harrie and Zadpoor, Amir A.},
  journal = {RSC Advances},
  volume = {6},
  pages = {20431--20436},
  year = {2016},
  doi = {10.1039/C6RA00295A}
}

@article{xu2015soft,
  title = {Soft Network Composite Materials with Deterministic and Bio-Inspired Designs},
  author = {Xu, Sheng and Yan, Zheng and Jang, Kyung-In and Huang, Wen and Fu, Haoran and Kim, Jeonghyun and Wei, Zhaoqian and Flavin, Matthew and McCracken, John and Wang, Rui and Badea, Alexandra and Liu, Yuhao and Xiao, Dongqing and Zhou, Guoying and Lee, Jungwoo and Chung, Hoon-Joon and Cheng, Huanyu and Ren, Wenguang and Banks, Anthony and Li, Xiuling and Paik, Uhn and Nuzzo, Ralph G. and Huang, Yonggang and Zhang, Yong-Wei and Rogers, John A.},
  journal = {Nature Communications},
  volume = {6},
  pages = {6566},
  year = {2015},
  doi = {10.1038/ncomms7566}
}

@article{ma2016nonlinear,
  title = {A Nonlinear Mechanics Model of Bio-Inspired Hierarchical Lattice Materials Consisting of Horseshoe Microstructures},
  author = {Ma, Qiang and Cheng, Huanyu and Jang, Kyung-In and Luan, Hailong and Hwang, Kyu-Jin and Rogers, John A. and Huang, Yonggang},
  journal = {Journal of the Mechanics and Physics of Solids},
  volume = {90},
  pages = {179--202},
  year = {2016},
  doi = {10.1016/j.jmps.2016.02.012}
}

@article{liu2016harnessing,
  title = {Harnessing Buckling to Design Architected Materials that Exhibit Effective Negative Swelling},
  author = {Liu, Jia and Gu, Tianyu and Shan, Sicong and Kang, Sung H. and Weaver, James C. and Bertoldi, Katia},
  journal = {Advanced Materials},
  volume = {28},
  number = {31},
  pages = {6619--6624},
  year = {2016},
  doi = {10.1002/adma.201600812}
}

@article{kang2013buckling,
  title = {Buckling-Induced Reversible Symmetry Breaking and Amplification of Chirality Using Supported Cellular Structures},
  author = {Kang, Sung H. and Shan, Sicong and Noorduin, Wim L. and Khan, Mughees and Aizenberg, Joanna and Bertoldi, Katia},
  journal = {Advanced Materials},
  volume = {25},
  number = {24},
  pages = {3380--3385},
  year = {2013},
  doi = {10.1002/adma.201300617}
}

@article{shan2015multistable,
  title = {Multistable Architected Materials for Trapping Elastic Strain Energy},
  author = {Shan, Sicong and Kang, Sung H. and Raney, Jordan R. and Wang, Pai and Fang, Lichen and Candido, Francisco and Lewis, Jennifer A. and Bertoldi, Katia},
  journal = {Advanced Materials},
  volume = {27},
  number = {29},
  pages = {4296--4301},
  year = {2015},
  doi = {10.1002/adma.201501708}
}

@article{guo2023nonorientable,
  title = {Non-Orientable Order and Non-Commutative Response in Frustrated Metamaterials},
  author = {Guo, Xiaofei and Guzm{\'a}n, Marcelo and Carpentier, David and Bartolo, Denis and Coulais, Corentin},
  journal = {Nature},
  volume = {618},
  pages = {506--512},
  year = {2023},
  doi = {10.1038/s41586-023-06022-7}
}

@article{udani2022taming,
  title = {Taming Geometric Frustration by Leveraging Structural Elasticity},
  author = {Udani, Janav P. and Arrieta, Andres F.},
  journal = {Materials \& Design},
  volume = {221},
  pages = {110809},
  year = {2022},
  doi = {10.1016/j.matdes.2022.110809}
}

@article{findeisen2017characteristics,
  title = {Characteristics of Mechanical Metamaterials Based on Buckling Elements},
  author = {Findeisen, Claudio and Hohe, J{\"o}rg and Kadic, Muamer and Gumbsch, Peter},
  journal = {Journal of the Mechanics and Physics of Solids},
  volume = {102},
  pages = {151--164},
  year = {2017},
  doi = {10.1016/j.jmps.2017.02.011}
}

@article{oppenheimer2015shapeable,
  title = {Shapeable Sheet without Plastic Deformation},
  author = {Oppenheimer, Naomi and Witten, Thomas A.},
  journal = {Physical Review E},
  volume = {92},
  pages = {052401},
  year = {2015},
  doi = {10.1103/PhysRevE.92.052401}
}

@article{ronceray2019range,
  title = {Range of Geometrical Frustration in Lattice Spin Models},
  author = {Ronceray, Pierre and Le Floch, Bruno},
  journal = {Physical Review E},
  volume = {100},
  pages = {052150},
  year = {2019},
  doi = {10.1103/PhysRevE.100.052150}
}

@article{ronellenfitsch2019inverse,
  title = {Inverse Design of Discrete Mechanical Metamaterials},
  author = {Ronellenfitsch, Henrik and Stoop, Norbert and Yu, Josephine and Forrow, Aden and Dunkel, J{\"o}rn},
  journal = {Physical Review Materials},
  volume = {3},
  pages = {095201},
  year = {2019},
  doi = {10.1103/PhysRevMaterials.3.095201}
}

@article{woodhouse2018autonomous,
  title = {Autonomous Actuation of Zero Modes in Mechanical Networks Far from Equilibrium},
  author = {Woodhouse, Francis G. and Ronellenfitsch, Henrik and Dunkel, J{\"o}rn},
  journal = {Physical Review Letters},
  volume = {121},
  pages = {178001},
  year = {2018},
  doi = {10.1103/PhysRevLett.121.178001}
}

@article{sharabani2022messy,
  title = {Messy or Ordered? Multiscale Mechanics Dictates Shape-Morphing of Two-Dimensional Networks Hierarchically Assembled of Responsive Microfibers},
  author = {Ziv Sharabani, Shiran and Edelstein-Pardo, Nicole and Molco, Maya and Bachar Schwartz, Noa and Morami, Maayan and Sivan, Avishai and Gendelman Rom, Yonatan and Evental, Roey and Flaxer, Eli and Sitt, Amit},
  journal = {Advanced Functional Materials},
  volume = {32},
  number = {19},
  pages = {2111471},
  year = {2022},
  doi = {10.1002/adfm.202111471}
}

@article{sharabani2024directional,
  title = {Directional Actuation and Phase Transition-Like Behavior in Anisotropic Networks of Responsive Microfibers},
  author = {Ziv Sharabani, Shiran and Livnat, Elad and Abuchalja, Maia and Haphiloni, Noa and Edelstein-Pardo, Nicole and Reuveni, Tomer and Molco, Maya and Sitt, Amit},
  journal = {Soft Matter},
  volume = {20},
  pages = {2301--2309},
  year = {2024},
  doi = {10.1039/D3SM01753B}
}

@article{keogh2025combinatorial,
  title = {Combinatorial Asymmetric Acoustic Metamaterials with Real-Time Programmability},
  author = {Keogh, Melanie R. and Bilal, Osama R.},
  journal = {Proceedings of the National Academy of Sciences},
  volume = {122},
  number = {48},
  pages = {e2502036122},
  year = {2025},
  doi = {10.1073/pnas.2502036122}
}

@article{kwakernaak2023counting,
  author  = {Kwakernaak, Lennard J. and van Hecke, Martin},
  title   = {Counting and Sequential Information Processing in Mechanical Metamaterials},
  journal = {Physical Review Letters},
  volume  = {130},
  pages   = {268204},
  year    = {2023},
  doi     = {10.1103/PhysRevLett.130.268204}
}

@article{meulblok2026transients,
  author  = {Meulblok, Colin M. and van Hecke, Martin},
  title   = {Transients and multiperiodic responses: a hierarchy of material bits},
  journal = {New Journal of Physics},
  volume  = {28},
  year    = {2026},
  doi     = {10.1088/1367-2630/ae45c7}
}

@article{meulblok2026path,
  author  = {Meulblok, Colin M. and Singh, Amitesh and Labousse, Matthieu and van Hecke, Martin},
  title   = {Path-Dependency and Emergent Computing under Vectorial Driving},
  journal = {Physical Review X},
  volume  = {16},
  pages   = {031023},
  year    = {2026},
  doi     = {10.1103/2pry-2kqq}
}

@article{dieleman2020jigsaw,
  title={Jigsaw puzzle design of pluripotent origami},
  author={Dieleman, Peter and Vasmel, Niek and Waitukaitis, Scott and van Hecke, Martin},
  journal={Nature Physics},
  volume={16},
  number={1},
  pages={63--68},
  year={2020},
  publisher={Nature Publishing Group UK London}
}

@article{pisanty2020putting,
  title={Putting a spin on metamaterials: Mechanical incompatibility as magnetic frustration},
  author={Pisanty, Ben and O{\u{g}}uz, Erdal C and Nisoli, Cristiano and Shokef, Yair},
  journal={SciPost Physics},
  volume={10},
  pages={136},
  year={2021}
}

@article{Ferreira2001DoubleGamma,
  author  = {Ferreira, Chelo and L{\'o}pez, Jos{\'e} L.},
  title   = {An Asymptotic Expansion of the Double Gamma Function},
  journal = {Journal of Approximation Theory},
  volume  = {111},
  number  = {2},
  pages   = {298--314},
  year    = {2001},
  doi     = {10.1006/jath.2001.3578}
}

@article{kuperberg1994symmetries,
  title={Symmetries of plane partitions and the permanent—determinant method},
  author={Kuperberg, Greg},
  journal={Journal of Combinatorial Theory, Series A},
  volume={68},
  number={1},
  pages={115--151},
  year={1994},
  publisher={Elsevier}
}

\end{document}